# Graphene nanowindows as a basis for creating mechanically robust nanohydroxyapatite bone lamellar scaffolds

Radovan Kukobat,[1,2]* Ranko Škrbić,[2,3]* Sanja Jovičić[2,4], Milka Matičić,[2] Aneta Stojmenovski,[2] Marko Bulajić,[5] Saša Marin,[5] Darko Bodroža,[1,6] Dragana Stević,[7] Suzana Gotovac Atlagić,[6] Nenad Ignjatović,[8] Fernando Vallejos-Burgos,[9] Elisa Mercadelli,[10] Jessica Ponti,[11] Francesco Sirio Fumagalli,[11] Andrea Valsesia[11]

[1]University of Banja Luka, Faculty of Technology, Department of Chemical Engineering and Technology, B.V Stepe Stepanovica 73, Banja Luka, the Republic of Srpska, Bosnia and Herzegovina

[2]University of Banja Luka, Faculty of Medicine, Centre for Biomedical Research, Save Mrkalja 16, Banja Luka, the Republic of Srpska, Bosnia and Herzegovina.

[3] University of Banja Luka, Faculty of Medicine, Department of Pharmacology, Toxicology and clinical Pharmacology, Save Mrkalja 16, Banja Luka, the Republic of Srpska, Bosnia and Herzegovina

[4] University of Banja Luka, Faculty of Medicine, Department of Histology and Embryology, Save Mrkalja 16, Banja Luka, the Republic of Srpska, Bosnia and Herzegovina

[5]University of Banja Luka, Faculty of Medicine, Department of oral surgery, Save Mrkalja 16, Banja Luka, the Republic of Srpska, Bosnia and Herzegovina

[6]University of Banja Luka, Faculty of Natural Sciences and Mathematics, Mladena Stojanovića 2, 78000 Banja Luka, the Republic of Srpska, Bosnia and Herzegovina

[7]University of Novi Sad, Faculty of Technology Novi Sad, Boulevard cara Lazara 1, Novi Sad, Serbia

[8]Institute of Technical Sciences of the Serbian Academy of Science and Arts, Knez Mihailova 35/IV, P.O. Box 377, 11000, Belgrade, Serbia

[9]Morgan Advanced Materials, 310 Innovation Blvd., Suite 250, State College, Pensilvania 16803, United States

[10]CNR-ISSMC (former ISTEC), Institute of Science, Technology and Sustainability for Ceramics, Via Granarolo 64, Faenza I-48018, Italy

[11]Directorate General Joint Research Centre, Directorate F - Health and Food - Technologies for Health (F.2), via Fermi 2749, 21027 ISPRA (VA), Italy

*Corresponding author's information

Tel.: +387 (0) 51 434 357; Fax.: +387 (0) 51 434 351, E-mail: radovan.kukobat@tf.unibl.org.

Tel.: +387 (0) 51 234 100, Fax.: +387 (0) 51 215 454, E-mail:ranko.skrbic@med.unibl.org

# Abstract

The role of graphene nanowindows in the nanohydroxyapatite bone scaffold preparation is important for the preparation of mechanically robust scaffolds and implementation in bone recovery. Here, we report a graphene-nanohydroxyapatite (G-nHAP) scaffold synthesized by the hydrothermal method, along with its formation mechanism and promising *in vivo* applications. The G-nHAP scaffold exhibits excellent mechanical strength comparable to that of cancellous bone. The nHAP was grown on 2D layers of graphene with nanowindows. The role of nanowindows was to attract electrostatically $Ca^{2+}$, $PO_4^{3+}$, and $OH^-$ precursors of nHAP, forming a layered structure of G-nHAP, in which nHAP nanorods of 23.8 nm in diameter and 84.7 nm in length were placed between graphene layers, as evidenced with molecular dynamic simulations. *In vivo* study showed mature and mineralized bone and osteoid after six weeks. This demonstrates the role of graphene nanowindows in the formation of nHAP scaffolds that are promising for future implementation in bone tissue regeneration.

**Key words:** Graphene-based scaffolds, Nanowindows, Bone formation, Nanohydroxyapatite, Mechanically robust.

## 1. Introduction

Graphene is an important material for the development of bone scaffolds because of its antimicrobial activity and mechanical strength.[1] Graphene has nanowindows[2–5] or defects that can serve as growing sites of nanohydroxyapatite (nHAP). Coupling the nHAP with graphene can lead to the firm structure of graphene-nHAP (G-nHAP) suitable for bone scaffolds.[6] Graphene monolayers have an intrinsic mechanical strength as high as ~130 GPa,[7] while the ultimate breaking strength of graphene oxide (GO) monolayers was reported to be ~80 GPa,[8] according to theoretical calculations. Typical bones possess a strength of 51 – 66 MPa in tension and 106 – 131 MPa in compression at the longitudinal axis.[9] This implies that graphene-like materials such as GO and graphene are promising for reinforcing the strength of hydroxyapatite (HAP) for the synthesis of bone scaffolds. GO is favorable because of the presence of oxygen functional groups that can attach the active centers of HAP during crystal growth, enabling stimulated and controlled growth of HAP on the graphene layers.[10] GO-based materials are promising for the future bone tissue regeneration scaffolds,[11] and thereby a number of works demonstrated a possibility for graphene-based HAP preparation solely and together with biocompatible polymers, improving the properties.[12–17]

GO with HAP and gelatin cryogels helped hip bone regeneration in diabetic rats through electrical stimulation, leading to improvements in compound muscle action potential with combinatorial therapy by 70 %.[18] GO with polyvinylidene fluoride-induced electrical charges in the scaffold contributed to the cell behavior, improving the tensile strength of the scaffold by 24.5 %.[19] Graphene oxide-hydroxyapatite/silk fibroin scaffold has enhanced porosity composed of semicrystalline roads in the graphene-like structure, stimulating the HAP growth mouse mesenchymal stem cell adhesion and proliferation, alkaline phosphatase secretion.[20] GO with nano HAP reduced hydrothermally showed efficient *in vivo* bone repair with pronounced bone cell growth and mineralization.[21] Structure of graphene-based HAP with the hierarchical porosity, having macropores of 90 – 130 μm that are favorable for cell migration and transport, and small micropores that can provide cell adhesion points, being promising for adhesion and proliferation of the stem cells during bone regeneration.[22] GO incorporated in gelatine HAP matrix by freeze drying method, gives superior mechanical stress as high as ~7 MPa after reinforcement with GO.[23] The composite of GO with HAP and gelatine was applied to 3D printing technology, giving a uniform 3D mesh that has a stress at the fracture point of ~10 MPa at a concentration of GO of 0.5 %.[24] 3D GO foam/ polydimethylsiloxane/ zinc silicate shows a decent mechanical strength, having a porous structure with demonstrated possibility for proliferation and osteogenic differentiation.[25] Also, 3D honeycomb porous carbon based on chitosan and hydroxyapatite has a certain mechanical strength with a significant advantage in promoting bone formation.[26] GO can significantly enhance the mechanical properties of HAP for bone scaffolds. Also, reduced GO (rGO) can enhance the mechanical properties of 3D printed scaffolds, reaching a compressive modulus of 0.43 MPa at the composite state of rGO, gelatin, and HAP.[27] Reinforcing the HAP could be achieved by in situ hydrothermal synthesis of GO-HAP that could ensure a good contact between GO and HAP. GO-HAP composite (12 %) was blended with copolymer poly-L-lactic acid, giving the scaffold enhanced mechanical strength as high as 22 MPa at the fracture point.[28] Graphene doped with nitrogen with HAP scaffold in agarose matrix shows significant enhancement of the mechanical strength, reaching ~120 MPa at the fracture point, promoting proliferation and viability of the mesenchymal stem cells according to *in vivo* studies.[29] Hydroxyapatite/poly- L -lactide composite biomaterial of a high mechanical strength similar to natural bone could be obtained by the hot pressing method.[30] Yet, GO with nHAP and polymers at different ratios does not provide suitable materials of high mechanical strength and efficiency for bone regeneration, leaving many challenges in the state-of-the-art synthesis of GO with nHAP.

New ways of synthesis of graphene-based nHAP porous scaffolds with high mechanical strength are to be explored. Synthesis of scaffolds on graphene oxide layers that possess nanowindows would be suitable for interlayered nHAP growth giving embedded graphene-based nHAP bone scaffolds. Moreover, surface charges at the edges of nanowindows,[3] owing to the presence of oxygen functional groups would be promising to attract the constituents on nHAP such as $Ca^{2+}$, $PO_4^{3-}$, and $OH^-$ during the crystal growth. This could enable the crystal growth of nHAP on the nanowindows, coating the nanowindows and creating interconnections between the layers. Aiming to achieve an ideal state of the G-nHAP with nHAP grown on nanowindows or defects of GO, we applied a hydrothermal method using calcium nitrate, phosphoric acid, and ammonium hydroxide as precursors. Experimental evidence of the G-nHAP successful synthesis was provided using electron microscopy observations, X-ray diffraction, Fourier transform infrared, and Raman spectroscopy, while molecular dynamic simulations revealed nHAP crystals located near nanowindows. G-nHAP was molded into short bars for demonstration of excellent mechanical properties under compression. *In vivo* experiments were conducted on Wistar rats, in which holes of 4 mm in diameter were made in the *lamina externa*, filled with G-nHAP, and after six weeks the growth of the mature bone was evidenced by histological tests. This study evidences the structure of nHAP grown near nanowindows on graphene layers and the successful implementation of G-nHAP in bone recovery.

## 2. Experimental Section

### 2.1. Materials and method

Graphene oxide was synthesized from Madagascar graphite following an improved Hummers method.[31] 1 g of graphite was mixed with 40 $cm^3$ of concentrated $H_2SO_4$ and 4.5 $cm^3$ of concentrated $H_3PO_4$ and 5 g of $KMnO_4$ followed by stirring at 250 rpm at a temperature of 35 – 40 °C. Exfoliated GO was washed five times with 5 % HCl and five times with water prior to use for the growth of nHAP. nHAP was synthesized following hydrothermal synthesis in an autoclave using analytical purity grade reagents of $Ca(NO_3)_2 \cdot 4H_2O$, $H_3PO_4$, and $NH_4OH$ purchased from Sigma Aldrich. $Ca(NO_3)_2 \cdot 4H_2O$ and $H_3PO_4$ were mixed to achieve the Ca/P ratio that is similar to the Ca/P ratio in human bone.[32] $NH_4OH$ was added to precipitate calcium and phosphate ions. GO dispersion of 0.1 wt.% was added to the reaction mixture and the concentrations were adjusted to be 0.005, 0.02, and 0.03 wt. %. The reaction mixture was aged at 80 °C for 24 h and treated in an autoclave at 180 °C for 2h (Figure S1).[33] After the hydrothermal synthesis, water was evaporated to yield a paste-like consistency that was further evaporated, and the powder was obtained (Figure S2). GO-nHAP was annealed at 60 °C for 3 h in a vacuum to yield reduced graphene oxide-nanohydroxiapatite (G-nHAP) (Figure S3). The nHAP with 0.02 wt. % of GO showed the highest surface area according to $N_2$ adsorption at 77 K (Figure S4) and was characterized in detail including *in vivo* studies.

### 2.2. *In vivo* studies

We used 6 male Wistar rats weighing 200 – 300 g for experimental *in vivo* studies. *In vivo* experiments were conducted according to ethical guidelines for the use of animals in research.[34] Two bone defects of 4 mm in diameter were made at the left and right side of the rat skull, serving as defects for filling the G-nHAP, nHAP, and Cerabone bone substitute (Botiss Biomaterials, Germany) as a reference. The control group without the scaffold served as a reference for comparison between filled and empty defects in the bone healing process. After the operation, the wounds were cleaned with a moistened gaze and plastered. The care of the rats was taken every day including cleaning the wounds and feeding. After 6 weeks, the rats were terminated and the skulls containing the G-nHAP, nHAP, Cerabone, and control group were stored and fixated in 4% formalin for 48 h. After decalcification in 3% nitric acid, tissue samples were processed using the automated Leica TP 1020 tissue processor and subsequently embedded in paraffin blocks with the Leica HistoCore Arcadia H device. The paraffin blocks were then sectioned using a Rotary 3003 pfm microtome into 4 μm thick slices, which were mounted onto standard glass slides from Epredia Netherlands B.V. The study was approved by the Ethical Committee (No. 18/4.116-1/246, January 17, 2024) at the Faculty of Medicine, University of Banja Luka.

Morphological parameters during the bone formation process were determined after staining the samples with hematoxylin and eosin, as well as those stained with Masson's trichrome using the automated Myreva. The volume density (Vv) of cells was estimated using QuPath software by detecting and quantifying cells within defined regions of interest (ROI) on H&E and Masson trichrome stained tissue sections. The number of detected cells (N) within each ROI was normalized to the estimated volume of the analyzed tissue, calculated as the product of the ROI area (A) and the section thickness (T). The volume density was calculated using the following formula:

$$V_v = \frac{N}{A \times T} \qquad (1)$$

where: $Vv$ = volume density of cells (cells/μm³), $N$ = number of detected cells, $A$ = area of the ROI (μm²), $T$ = section thickness (μm). The numerical areal densities ($N_A$) were determined as follows:

$$N_A = \frac{N}{A} \qquad (2)$$

The microphotographs were taken at the magnification of 400x, while the measurements were performed in 10 visual fields.

## 2.3. Materials characterization

Materials characterization was conducted to understand the structure, composition, and properties of the G-HAP. Surface area and porosity parameters of nHAP and G-HAP were calculated from $N_2$ adsorption isotherms at 77 K using a volumetric gas adsorption analyzer (3P INSTRUMENTS). Pore size distribution was calculated using the Barrett-Joyner-Halenda (BJH) method.[35] Fourier transform infrared spectrometer (FTIR, BRUKER, Tensor 27, Germany) was applied to examine the surface functional groups and the possible interactions between reduced graphene oxide and HAP. Raman spectrometry (Renishaw, UK) served for examination of the crystallinity of GO, G-HAP, and the structure of nHAP. X-ray diffractometer (Bruker , Cu K-alpha, $\lambda = 1.5406$ Å, X-ray lamp power 1600W, I = 40 mA and V = 40 kV) was used to examine the crystallinity of nHAP and G-nHAP and the crystal sizes were evaluated by Scherrer equation. Transmission electron microscopy (JEOL JEM-2100, 120 kV) served for observation of nHAP and G-nHAP. The crystal sizes from the TEM images were evaluated using ImageJ.[36] Observation of the bulk phase was conducted after embedding the G-nHAP and nHAP in epoxy resin; the epoxy resin was sliced prior to the bulk phase observation. Scanning electron microscope coupled with elemental dispersive spectroscopy (EDS) (SEM, JEOL JSM-7800F FESEM) served for the surface observation and estimation of the Ca/P ratio in the scaffolds. Mechanical strength (Shimadzu AGS-X 10 KN) in compression mode at 1.0 mm min⁻¹ was conducted to understand the mechanical properties of the nHAP and G-nHAP. The stress-strain curves were acquired for nHAP and G-HAP, and the corresponding fracture and Young's modulus were evaluated. Optical microscopy observations of the rat skull were performed using a binocular Leica DM 6000 microscope, equipped with a Leica DFC310FX camera.

### 2.4. Computational details

We used a Large-scale Atomic/Molecular Massively Parallel Simulator (LAMMPS) to simulate the local structure of the HAP crystal with reduced graphene oxide. Simulations were conducted in a water medium using the TIP4P model (Table S1).[37] The temperature was set to be at 300 K using a Nose Hoover thermostat. The simulations were conducted in the NVT ensemble. HAP unit cell ($Ca_{10}(PO_4)_6(OH)_2$) with the dimensions of a = 0.9412 nm, b = 0.9412 nm, and c = 0.6853 nm was extracted from the crystallography data.[38–40] We prepared the model of GO with a nanowindow of a size of ~1 nm in diameter because it was typical nanowindow size by this method as reported in our previous work.[5] The edges were functionalized with carboxyl, hydroxyl groups, and hydrogen atoms. Two-unit cell crystals of HAP were placed between the layers of functionalized GO and simulations were conducted in water at the density of 1000 kg m⁻³. GO model was relaxed using ReaxFF[41] and combined with two unit cells of HAP in a simulation box of $4 \times 8 \times 5$ nm$^3$ with periodic boundary conditions. We used Lennard-Jones simulation parameters for the atoms in the system (see Table S1).

## 3. Results and discussion

### 3.1. Mechanically robust G-nHAP for bone recovery

The G-nHAP was applied to the rat mouse skull, referring to the standard Cerabone bone substitute. Two defects on the left and right sides of the skill were made on the lamina externa and diploë, while the lamina interna remained intact (Figure 1a). The G-nHAP and Cerabone as a reference were filled into the skull defects, showing the bone recovery after six weeks, as observed macroscopically (Figure 1b). The newly formed bone completely covered the holes on the skull filled with G-nHAP and a standard Cerabon bone substitute, indicating successful bone healing, being promising for the future implementation of the G-nHAP as a bone substitute and scaffold.

G-nHAP scaffolds were prepared by hydrothermal *in situ* synthesis method by introducing the diluted GO dispersion at the intermediate synthesis step (Figure S1a-c). This method should enable the synthesis of nHAP embedded between the layers of GO. Further, the GO-nHAP was thermally treated in the colloidal dispersion state, leading to hydrothermal reduction[42] of GO to reduced GO (denoted as G) due to instability and metastability of GO layers that were prone to spontaneous changes and reduction through $\pi - \pi^*$ transitions[43]. A dark G-nHAP scaffold with a paste-like consistency that could be molded and shaped was suitable for future manipulation and fillings into the bone defects (Figure 1c). GO-nHAP can be dried at room temperature for 12 h and ground, giving the powder-like consistency (Figure S2). The change in the G-nHAP from paste to powder form is reversible, suggesting that after mixing the G-nHAP with water, we could obtain the paste form again, being suitable for processing.

The high mechanical strength of the G-nHAP scaffold was demonstrated by compression measurements of the G-nHAP with the reference of the nHAP scaffold. The PTFE mold was filled with the nHAP and G-nHAP (Figure 1c), and solidified after drying at 25 °C for 24 h at an ambient condition, followed by a vacuum treatment at 250 °C for 2 h to remove remained water and additionally reduce GO, giving strong nHAP and G-nHAP scaffolds (Figure 1d). The scaffold is composed of graphene-like layers between the grains of the nHAP crystals as shown by a plausible model (insert in Figure 1e). The cylindrical scaffolds were tested by slow compression, showing the mechanical strength of the nHAP and G-nHAP (Figure 1f). Ultimate strength reached 12.94 ± 1.47 MPa for the G-nHAP and 0.086 ± 0.091 MPa for nHAP, while Young's modulus of elasticity reached 161.83 ± 48.37 MPa for G-nHAP and 39.46 ± 43.85 MPa for nHAP (Figure 1g,h, Table 1). The mechanical strength and elasticity of the G-nHAP scaffold are significantly greater than that of the nHAP due to the presence of the graphene-like structure embedded between the nHAP crystalline grains. Embedding the G-nHAP structure during the growth of the bone could contribute to an enhancement of the mechanical strength of the recovered bones.[22] An enhancement of the mechanical strength after embedding the graphene-like structure between nHAP occurs due to the strong interfacial strength between graphene and nHAP.[20] The G-nHAP bone scaffold has a high compressive mechanical strength that is comparable to the cancellous bone with a strength of 2 – 38 MPa.[44] The G-nHAP has similar mechanical strength as the cancellous bone, being suitable for avoiding the stress shielding effect. Nevertheless, the mechanical strength of the scaffold treated at 60 °C is strong enough to support the efficient growth of new bone, as demonstrated by pathohistological tests. Furthermore, the G-nHAP treated at 100 and 200 °C has an ultimate strength of nearly 2 MPa, being almost in the range of cancellous bone (Figure S5, Table S2). Thus, tuning the mechanical strength of the G-nHAP is possible by oxidation of graphene at different temperatures. The stress decreased after thermal treatment at 100 and 200 °C due to additional oxidation of graphene and a decrease of crystalline $sp^2$ carbon atoms that are responsible for high mechanical strength. Graphene possesses oxygen functional groups on its surface that were activated at higher temperatures, leading to a decrease in the ultimate strength from 12.94 ± 1.47 MPa at 60 °C to 1.57 ± 0.71 MPa at 100 °C, and 1.12 ± 0.37 MPa at 200 °C (Table S2). Also, the other mechanical property parameters decreased after the thermal treatment. This decrease in the mechanical property parameters is attributed to the oxidation of graphene in the presence of intrinsic oxygen. The strong interfacial interactions could contribute to the compact bone scaffold formation and its integration with natural bone.

**Table 1.** Mechanical property parameters including Young's modulus, Yield strength, and ultimate strength of the G-nHAP and nHAP scaffolds. The data were averaged using statistical average and standard deviation.

| Scaffold | Young's modulus (MPa) | Ultimate strength (MPa) | Yield strength (MPa) |
| --- | --- | --- | --- |
| G-nHAP | 161.83 ± 48.37 | 12.94 ± 1.47 | 11.49 ± 0.72 |
| nHAP | 39.46 ± 43.85 | 0.086 ± 0.091 | 0.059 ± 0.054 |

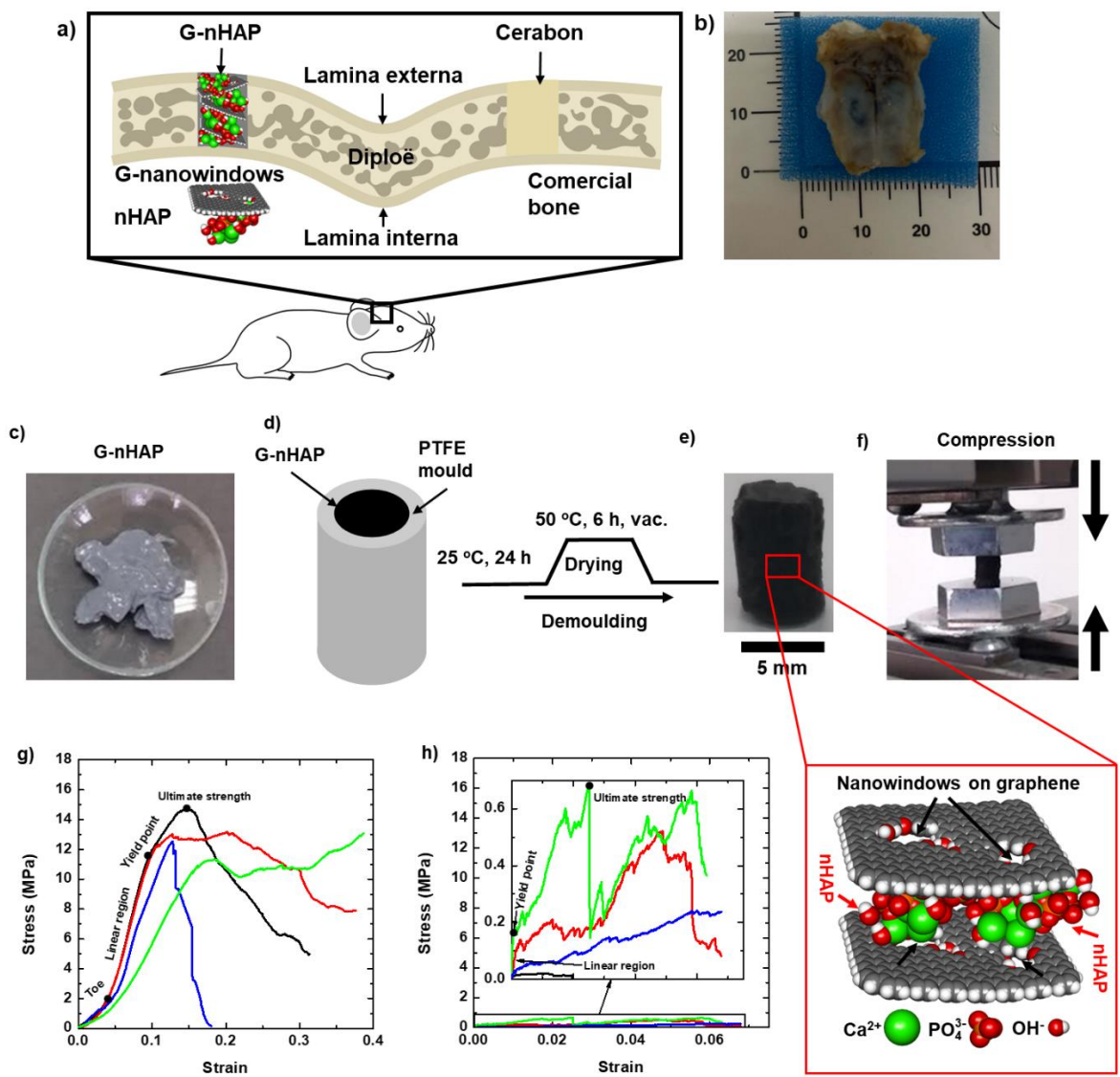

**Figure 1.** G-nHAP as a mechanically robust bone scaffold for bone recovery. (a) Skull with lamina interna, lamina externa, and diploë, showing two holes with G-nHAP and Cerabone bone substitute. (b) The calvaria shows the recovered bone at the place of the G-nHAP and Cerabone bone substitute. (c) G-nHAP after

water evaporation, having a paste-like morphology. (d) Molding a G-nHAP in a PTFE mold. (e) G-HAP after demoulding and drying at 25 °C for 24 h and 95 °C for 2h. An insert shows a microscopic model of a hybrid structure of graphene and nHAP. (f) Compression of the G-nHAP using a tensile strength machine at a compression mode at the speed of 1.0 mm min$^{-1}$. (g) Compressive stress vs. strain curve for G-nHAP. (h) Compressive stress vs. strain curve for nHAP.

## 3.2. Microstructure of the G-nHAP scaffolds

Observations of the microstructure of the scaffolds are crucial for the future understanding of their mechanical strength and functionality for *in vivo* (preclinical and clinical) applications. We examined the structure of the scaffolds by TEM and SEM observations (Figure 2,3). Interestingly the nHAP in the scaffold has a rod-like structure (Figure 2). Overall, the nHAP structure solay and nHAP with the graphene-like layers have similar rod-like shapes and morphology (Figure 2a–b,e– f). However, the dimensions (including the lengths and diameters) of the nHAP and G-nHAP are different, suggesting an effect of the graphene-like layers during the synthesis process. The mean length of the nHAP rods of 103.1 nm is longer than that of the G-nHAP rods of 84.7 nm, while the mean diameter of the nHAP rods of 16.6 nm is shorter than that of the G-nHAP rods of 23.8 nm (Figure 2c–d,g–h). The G-nHAP and nHAP have a needle-like structure that is similar to the nHAP structure in the lamellar bone structure.[45] The sizes and shapes of the G-nHAP and nHAP are similar to those reported in the literature, ranging from 50 – 100 × 30 – 80 nm.[46–49] The nHAP crystals play a role in the formation of the lamellar bone structure, connecting the collagen triple helix structure with osteoblasts forming the robust bone lamellar structure.[45]

The size differences between nHAP and G-nHAP were likely affected by an interfacial growth mechanism between graphene with nanowindows and nHAP. The mechanism of nHAP growth occurs through the nucleation and growth process.[50] Rod shape structure was formed due to the hexagonal crystallographic nature of hydroxyapatite which has a group space P6$_3$/m and the dimensions in x-, the y-axis of 0.9432 nm (x = y), and z-axis of 0.6881 nm at an angle of γ = 120°.[51] Owing to the slight solubility of calcium and phosphate ions in the growth medium and the small concentration of those ions, rod growth occurs slowly along the z-axis, forming the hexagonal crystal structure.[52] However, the rod growth in length was limited due to the presence of graphene layers with nanowindows in the growth medium. These graphene layers with nanowindows suppressed the growth of the nHAP rods in length while leading to an increase in the diameter of the nHAP. The growth mechanism is governed by the interface of graphene and nHAP, in which the nHAP precursor ions of $Ca^{2+}$, $PO_4^{3-}$, $NH_4^+$, and $OH^-$ were located near graphene nanowindows (section 3.4). Ca and P are the major elements uniformly distributed across the area of the nHAP and G-nHAP scaffolds (Figure S6). The Ca/P ratio of the scaffolds before and after introducing graphene was $1.89 \pm 0.05$ for nHAP and $1.84 \pm 0.04$ for G-nHAP (Table S3), being similar to that of the natural bone[53]. The interfacial crystal growth between nHAP precursors such as Ca/P and graphene-nanowindows contribute to a decrease in the length of nHAP by 21 % and an increase in diameter by 42 %. This indicates that the graphene layers were barriers to the nHAP growth, while on the other hand the graphene-like layers wrapped[54] the nHAP leading to an increase in the diameter. Thus, the nHAP and G-nHAP have rod-like structures with the nHAP near the surface of graphene nanowindows.

The pore structure and porosity of the microstructural G-nHAP were observed after cutting the G-nHAP in the silicon resin (Figure 3). The microstructure of the nHAP and G-nHAP was investigated by TEM observations after solidification with the polymer resin. After cutting the nHAP and G-nHAP, we observed porosity at the cross-sectional area. Indeed, the presence of circular pores of 0.1 to 1 µm can be observed (Figure 3 and Figure S7). These pores are relatively large, having a nonuniform size distribution, suggesting the presence of a hierarchical pore structure.[22] Existence of porosity in the bulk state of the nHAP is very important for bone recovery. In addition, the pore size distribution of the G-nHAP scaffolds was evaluated from the $N_2$ adsorption isotherms at 77 K (Figure S4a). The $N_2$ adsorption isotherms show low nitrogen adsorption uptake at the low-pressure region of $< 10^{-3}$, suggesting that the micropores are not dominant in the scaffold. The pore volume suggests that the G-nHAP scaffold at 0.02 wt.% of graphene has the highest porosity (Figure S4b). The pore size distribution of the scaffolds shows prominent pores with radii of 1.30 and 1.80 nm. Macropores are important for cell proliferation and growth because the cell sizes can fit large macropores. In contrast, micropores are important for the adsorption of proteins and can stimulate the interfacial interactions between the G-nHAP scaffold and cells.[55] The cell proliferation and osteogenic differentiation can be stimulated by the high porosity of the scaffolds, enabling efficient bone recovery.[56] The porosity of the scaffolds varies from uncontrolled pore sizes and shapes to the controlled shapes from the nano-, micro- to centi-meter orders of magnitude.[22,23,28] Graphene-like layers surrounding the nHAP rods during the crystal growth contribute to the excellent mechanical strength of the bone scaffolds, preserving high porosity that is important for cell proliferation.

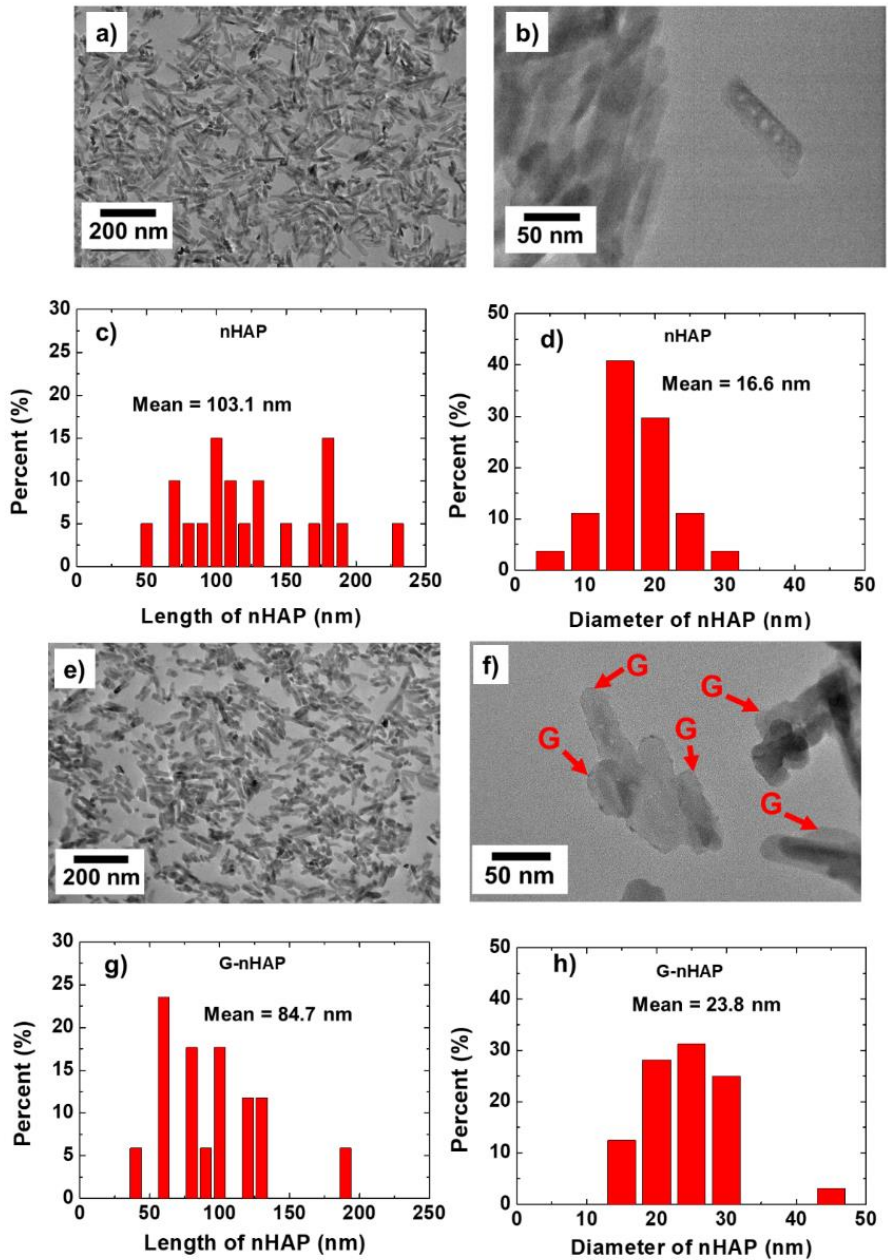

**Figure 2.** TEM micrographs of nHAP and G-nHAP. (a,b) Needle-like structures of the nHAP crystals. (c) Length and (d) diameter of the nHAP crystals. (e,f) Needle-like structures of G-nHAP. The "G" in red

denotes the layers of reduced graphene oxide covering the needle-like structures of G-nHAP. (g) Length and (h) diameter of G-nHAP needles.

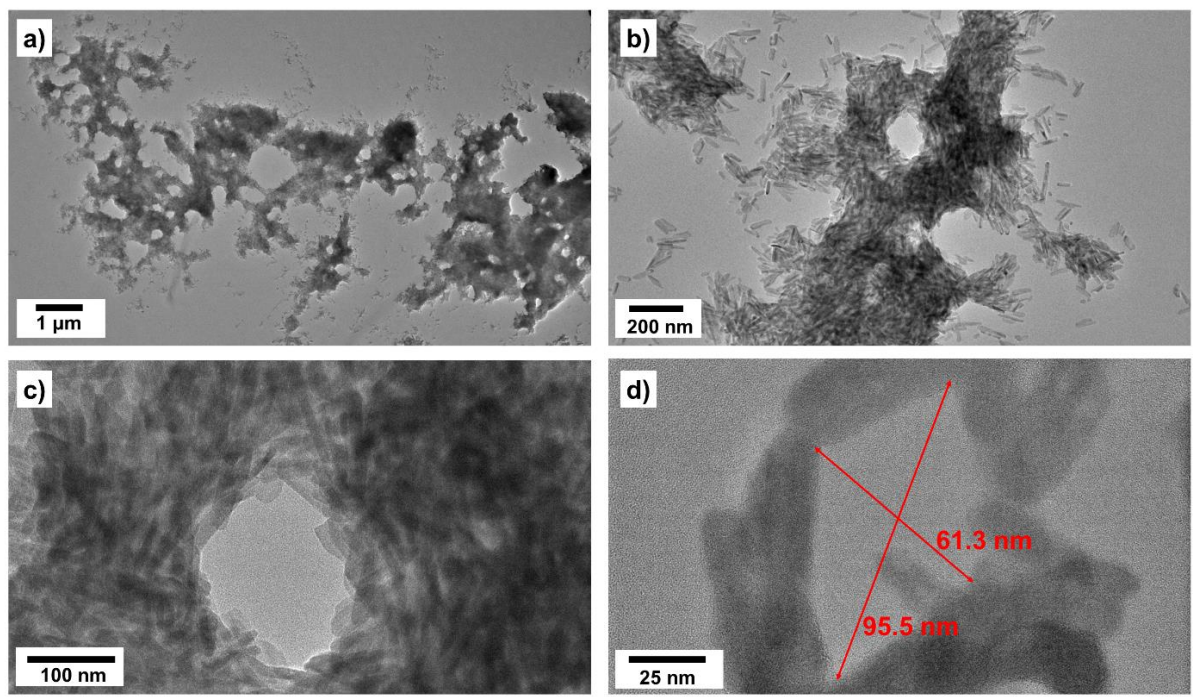

**Figure 3.** TEM micrographs of the G-nHAP observed at the slices after cutting the silicon resin. (a) Low magnification micrographs, showing the hierarchical pores. (b) Local pore structure, showing the presence of several pores. (c) Local pore structure formed of the G-nHAP rods. (d) The pore structure formed from the G-nHAP rods.

### 3.3. Structure of the scaffolds

The structure of the scaffold was investigated to understand the role of a graphene-like structure integrated with the nHAP rods using XRD (Figure 4a). The XRD patterns show the peaks characteristic for the hexagonal crystal structure of hydroxyapatite, according to the standard hydroxyapatite JCPDS (9-0432).[57] All peaks from the crystallographic faces including (002), (210), (211), (112), (300), (301), (310), (222), (213), (004) and (510) shifted after growth of nHAP with graphene-like structure indicating strong interactions between the crystal faces and graphene with nanowindows. The crystallographic sizes (Table 2) of nHAP calculated from the Scherrer[58] equation show altering the sizes at each crystal face in presence of graphene-like structure. These changes come from the interference of the graphene layers with nanowindows and limited crystal growth. The most prominent decrease in the crystal size by 3.42 nm occurred at the crystallographic face (004), while an increase by 7.75 nm occurred at the crystal face (002) (Table 1). On the other hand, the decrease in the crystal sizes at the crystallographic face (211) was as small as 0.93 nm. An alternate decrease and increase in the crystallographic faces occurs due to the presence of the 2D graphene-like barriers to the growth of the nHAP, leading to an uneven crystal growth. The crystal size of nHAP and G-nHAP reaches 19.67 and 26.42 nm at the crystalographic face (002). These crystal sizes almost coincide with the diameters of nHAP and G-nHAP obtained from the TEM micrographs (Figure 2d,h). This indicates that the graphene layers tightly wrapped[54] the nHAP rod, while the crystal growth occurred along the z-axis, giving elongated nHAP rods.

Raman spectra were used to study the interactions between the graphene-like layers and the nHAP rods. The nHAP rods show prominent bands from the $PO_4^{3-}$ anion at the wavenumbers of 432, 592, 963, and 1054 $cm^{-1}$,[59] while graphene-like structure shows the characteristic D-band[60] at 1356 $cm^{-1}$ and G-band[60] at 1600 $cm^{-1}$ (Figure 4b). The bands at the wavenumbers of 432 and 592 $cm^{-1}$ come from the triply (P-O-P) degenerate bending mode[61] of the $PO_4^{3-}$ anion of $Ca_3(PO_4)_2$ that is one of the constituents of the nHAP rods. The bands at the wavenumbers of 963 and 1054 $cm^{-1}$ come from symmetric and asymmetric stretching of the P-O group of the $PO_4^{3-}$ anion. The band shifts of the G-nHAP referring to the nHAP for symmetric and asymmetric stretching of the $PO_4^{3-}$ anions occurred slightly to the lower frequency region, while the intensity decreased significantly. An intensity decrease occurred due to the possible covering of the nHAP rods with the graphene-like crystals suppressing active vibrations of the nHAP constituents. On the other hand, the D-band which comes from out-of-plane vibrations of defective $sp^3$ carbon atoms at the edges of nanowindows increased and shifted to the lower frequency region by 43.73 $cm^{-1}$, while the G-band which comes from the $sp^2$ carbon atoms from in-plane vibrations shifted to the higher frequency region by 11.37 $cm^{-1}$ after addition of the graphene-like structure to nHAP (G-nHAP). An increase in the D/G band intensity from 0.94 for graphene-like structure to 0.99 for G-nHAP and the band shift suggests that the nanowindows of graphene enlarged in contact with nHAP in addition to the strong interactions at the sites of graphene with nanowindow and the nHAP rods. The higher frequency shift of the G-band in the presence of graphene indicates the charge transfer interactions from the nHAP rods to graphene. Thus, the Raman spectra of nHAP and graphene-like structure suggested important information about defective graphene with nanowindows, the structure of the nHAP rods, and their interactions.

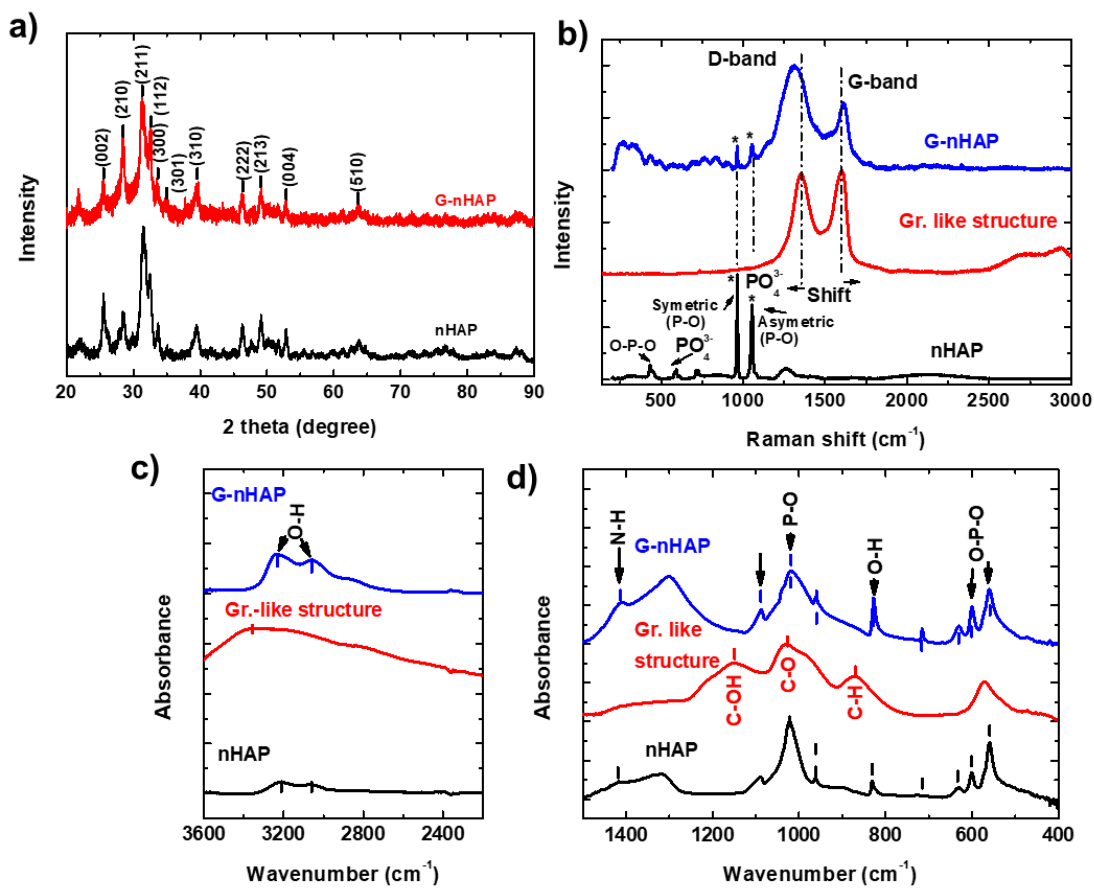

**Figure 4.** Structural analysis of the G-nHAP and nHAP bone scaffolds. (a) XRD patterns. (b) Raman spectra recorded at the laser wavelength of 532 nm. (c) FTIR spectra at the fingerprint region and (d) a narrow-spectra at $3600 - 2300$ cm$^{-1}$.

The structure and functionality of nHAP and G-nHAP were examined by FTIR analysis (Figure 4c,d). The broad band at the frequency region of $3600 - 2400$ cm$^{-1}$ is assigned to $O - H$ stretching vibrations of adsorbed $H_2O$ (Figure 4c).[62] The G-nHAP and nHAP rods have two distinctive bands at 3056 cm$^{-1}$ from four tetrahedrally arranged hydrogen bond subpopulations of $H_2O$ molecules, while the band at 3231 cm$^{-1}$ comes from O-H groups of $H_2O$ connected with the distorted hydrogen bonds.[63] A broad band at 3365 cm$^{-1}$ comes from physisorbed $H_2O$ bound through hydrogen bonding in a multilayered space of graphene-like structure.[64] The nHAP rods have the characteristic bands at 1019 cm$^{-1}$ from P-O stretching vibrations in $PO_4^{3-}$ anion, at 557 and 600

cm$^{-1}$ from O-P-O symmetric and asymmetric vibrations in PO$_4^{3-}$ anion, while the band at 1390 cm$^{-1}$ comes from N-H that remained trapped inside the structure during the dissolution process in the synthesis.[65] Graphene-like structure with the nanowindows has bands at 870 cm$^{-1}$ from C-H out of plane vibrations, at 1025 cm$^{-1}$ from C-O, and at 1149 cm$^{-1}$ from from C-OH vibrations.[66] The vibrational frequency band of the G-nHAP at 1019 cm$^{-1}$ from the P-O functional groups broadened and shifted by 2.33 cm$^{-1}$ to the lower frequency region, suggesting the presence of the interactions between graphene and nHAP. This suggests that the graphene-like structure was attached to the nHAP via PO$_4^{3-}$ groups. Thus, the crystal growth was restricted at the sides where graphene layers were attached to the nHAP crystal, as suggested by XRD analysis of the nHAP rod sizes.

**Table 2.** Crystallographic sizes of the nHAP and G-nHAP evaluated from the Scherrer equation. The signs (+) and (–) in the crystal size difference column indicate an increase and a decrease in the nHAP crystal after the addition of graphene.

| Crystal face | Crystal size (nm) | | Crystal size difference (nm) |
|---|---|---|---|
| | nHAP | G-nHAP | |
| (004) | 20.12 | 16.70 | – 3.42 |
| (213) | 18.66 | 19.80 | + 1.14 |
| (222) | 15.74 | 14.54 | – 1.20 |
| (310) | 9.10 | 10.42 | + 1.32 |
| (300) | 19.62 | 17.74 | – 1.88 |
| (112) | 16.72 | 18.39 | + 1.67 |
| (211) | 10.19 | 9.26 | – 0.93 |
| (210) | 9.20 | 15.70 | + 6.50 |
| (002) | 19.67 | 27.42 | + 7.75 |

### 3.4. Model of the scaffolds and the role of graphene nanowindows

The microstructure of the scaffolds was investigated using MD simulations to understand how the nanocrystals of the nHAP rods were oriented against the graphene-like structural layers. The graphene-like structure with the nanowindows of 0.49 and 1.05 nm (Figure 5a) in diameter and terminated edges were used to calculate the potential energy of graphene layers and observe its orientation against the nHAP crystals. The small nHAP crystal consisted of $Ca^{2+}$, $PO_4^{3-}$, $OH^-$, and the graphene with the nanowindows were simulated (Figure 5a). Two triangular subcells (Figure 5b) of the nHAP were in a cell with a graphene layer and simulated with the force field in water at room temperature. The structure stabilized after the simulation run of nearly 100 ps, reaching the minimal potential energy (Figure 5c). The distance between the nHAP subcell reached the minimum of 0.6 nm in water medium (Figure 5d). The nHAP subcell moved to the nanowindow site and the $PO_4^{3-}$, while the $Ca^{2+}$ cation was oriented to the nanowindow (Figure 5e). This suggests that the nucleation and the crystal growth should occur near the nanowindow layers rather than on the crystalline graphene plains, supporting our assumption about graphene orientation to nHAP from FTIR spectral analysis. Considering the affinity of the nHAP crystals for the graphene-like structures with nanowindows, the crystal growth could be affected by the presence of graphene, affecting the sizes of the crystals (Table 1). The small microcrystals could form large microporous crystals by building together through the short- and long-range interactions. These large crystals could form the macroporous structures that are available for cell proliferation and future uses as bone scaffolds.

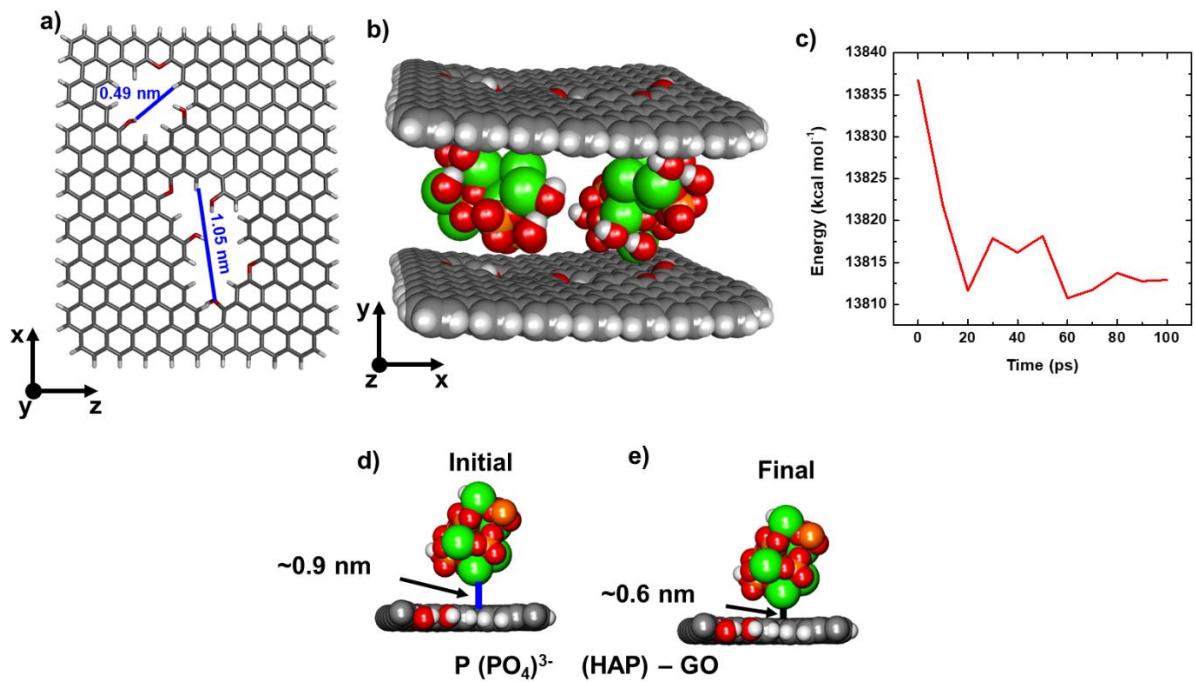

**Figure 5.** MD simulation of G-nHAP system. (a) Graphene-like structure with nanowindows. (b) A model of the G-nHAP system. (c) The potential energy of graphene at the interface of nHAP between graphene layers. (d) Cross-sectional view of the initial position of the nHAP crystal on the top of the graphene layer with nanowindow. (e) Cross-sectional view of the nHAP layer after stabilization of the system.

### 3.5. *In vivo* bone formation

The process of bone tissue healing and new bone formation is a multi-phase process involving the interaction of inflammatory cells, osteoprogenitor cells, and osteoblasts with extracellular molecules. Unlike other tissues, bone healing does not result in scar formation; instead, the newly formed bone tissue is indistinguishable morphologically and functionally from the surrounding bone. Microscopically, Hematoxin-eosin stained micrographs show lamina interna and trabeculae, together with the graphene layers that are responsible for the bone hardness (Figure 6a, a$_1$). Masson trichrome stained micrographs show the presence of mature bone, mineralized bone, osteoids, osteocytes, and osteoprogenitor cells (Figure 6b, b$_1$).

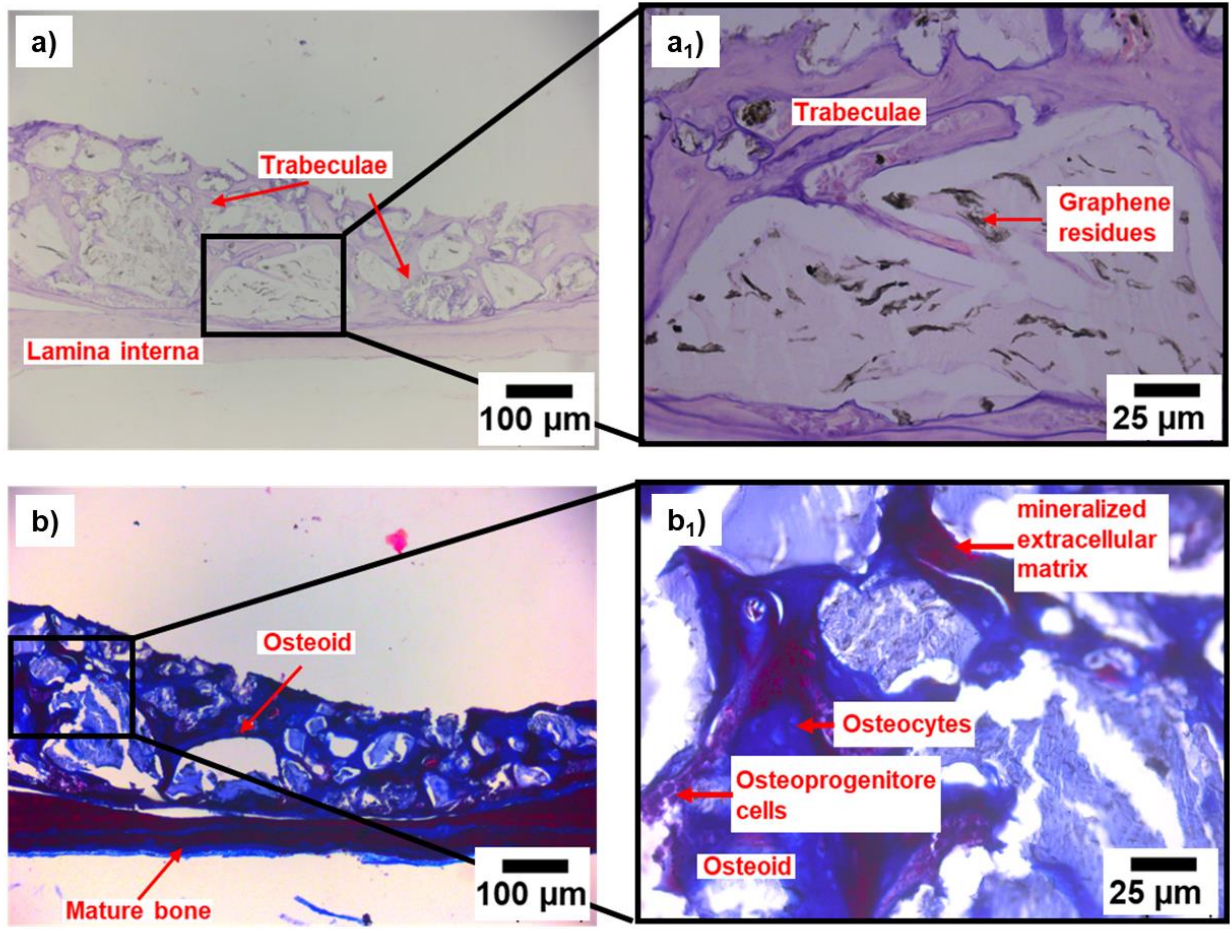

**Figure 6.** *In vivo* study of the G-nHAP scaffold with reference to Cerabone bone substitute. (a) and (a₁) Hemotoxin-eosin stained micrographs at the magnification of 100 and 400 times, respectively. (b) and (b₁) Masson trichrome stained micrographs at the magnification of 100 and 400 times, respectively.

The volume and areal densities show clear evidence of the effect of the G-nHAP, with references to the control group (empty defect) and nHAP (Table S4). The volume density of osteocytes and osteoblasts of the G-nHAP filled defects was $\sim$ ~60% and $\sim$ ~40% greater than that of the control group, respectively (Figure S8). Similarly, the numerical areal density of osteocytes and osteoblasts filled with the G-nHAP was $\sim$ ~28% and 40% greater than that of the control group, respectively. Interestingly, the volume and areal densities of osteocytes and osteoblasts of the G-nHAP filled defects were significantly higher than that of the nHAP, suggesting that graphene positively affects bone cell growth and differentiation. G-nHAP has a higher surface area than nHAP, giving more sites for cell adhesion, nutrient exchange, and adsorption of growth factors and proteins that are essential for osteogenesis.[67] In addition, the hydrophilicity of graphene is favorable for cell attachment and spreading.[68] The micrographs show preserved structures of the inner lamina, consisting of a mineralized bone matrix (Figure S8). The inner lamina is covered by an unmineralized bone matrix (osteoid), indicating areas of newly formed bone. The section reveals abundant connective tissue richly vascularized and populated with osteoprogenitor cells that will give rise to new bone trabeculae (future trabecular bone). Interestingly, the volume of newly formed bone, as well as the foundation for future trabecular bone is significantly greater for the G-nHAP scaffold (Figure S8b, $b_1$) than that of the empty bone defect (Figure S8a, $a_1$). Thus, the value of the G-nHAP is significant in the new bone formation, being a promising scaffold.

The analysis of the G-nHAP filled defects suggests the presence of trabecular bone and osteoprogenitor cells that are precursors for osteoblasts, being responsible for bone formation (Figure S9). The micrograph reveals newly formed trabecular bone closely associated with dense connective tissue. This connective tissue is richly vascularized and populated with numerous blood vessels, fibroblasts, and osteoprogenitor cells. Notably, graphene residues (arrows) are detected exclusively within the connective tissue and are absent from the newly formed bone matrix, indicating no incorporation into the bone structure. Osteoprogenitor cells are stem cells found in the periosteum (the outer layer of the bone) and endosteum (the inner layer of the bone). They are activated during bone regeneration, particularly after injuries. These cells are essential for bone growth and repair due to their ability to proliferate and differentiate. Osteoprogenitor cells differentiate into osteoblasts, which are responsible for producing osteoid, the unmineralized organic component of the bone matrix that serves as a scaffold for mineralization. Its flexibility allows bones to resist fractures before it is hardened by the deposition of minerals. Mineralization of the osteoid leads to the formation of mineralized bone. This process involves the deposition of calcium and phosphate, primarily in the form of hydroxyapatite crystals. Mineralization enables bones to withstand mechanical forces and store vital minerals, such as calcium and phosphorus, for metabolic processes in the body. After osteoblasts produce osteoids, they remain trapped within the bone matrix, which leads to their differentiation into osteocytes. Osteocytes are mature bone cells located within small spaces called lacunae. They are interconnected through long cytoplasmic extensions that run through microscopic channels called canaliculi. Osteocytes maintain the bone matrix by regulating mineral homeostasis. They act as mechanosensors, detecting mechanical stress and signaling for bone remodeling when necessary. Mature bone, also known as lamellar bone, is the final, fully developed form of bone with a highly organized structure. It consists of collagen fibers arranged in parallel layers (lamellae), which provide strength and resistance to stress. It replaces primary bone during remodeling. Mature bone offers structural support, durability, and the ability to withstand mechanical forces.

The G-nHAP scaffold is stable, according to *in vivo* experiments. The pH during bone healing was estimated to be in the range of 6.8 – 7.4, according to direct measurements using a high-resolution fiber optic pH microsensor.[69] An almost neutral pH is favorable for preserving the G-nHAP scaffold during the healing process. The constituents of the G-nHAP are stable at neutral pH, as each component, including graphene and nHAP,[70,71] exhibits stability under these conditions. The morphology of bone-healed defects after six weeks shows a preserved, non-degraded G-nHAP scaffold (Figure S10).

 The control group shows fewer and smaller ossification centers (shown by arrows) (Figure 7a,$a_1$) than the defect filled by G-nHAP (Figure 7b,$b_1$). In addition, the control the defect filled with the G-nHAP shows more pronounced vascularization than the control group with an empty defect. The mature bone quantity is more significant at the defect filled with the G-nHAP than that of the empty defects (Figure 7$a_1$,$b_1$). Blue staining indicates a non-mineralized bone matrix (osteoid), while red staining marks a mineralized bone matrix. Numerous osteoblasts (white arrows) and osteocytes (black arrows) are visible within the bone matrix.

The newly grown bone in the defect filled with the G-nHAP (Figure 7b,$b_1$) shows significantly increased ossification activity and trabecular bone formation in comparison with the control group (Figure 7a,$a_1$). Thus, enhanced vascular networks and osteoprogenitor cell density are evident at the defect filled with the G-nHAP, showing the promising application of this scaffold.

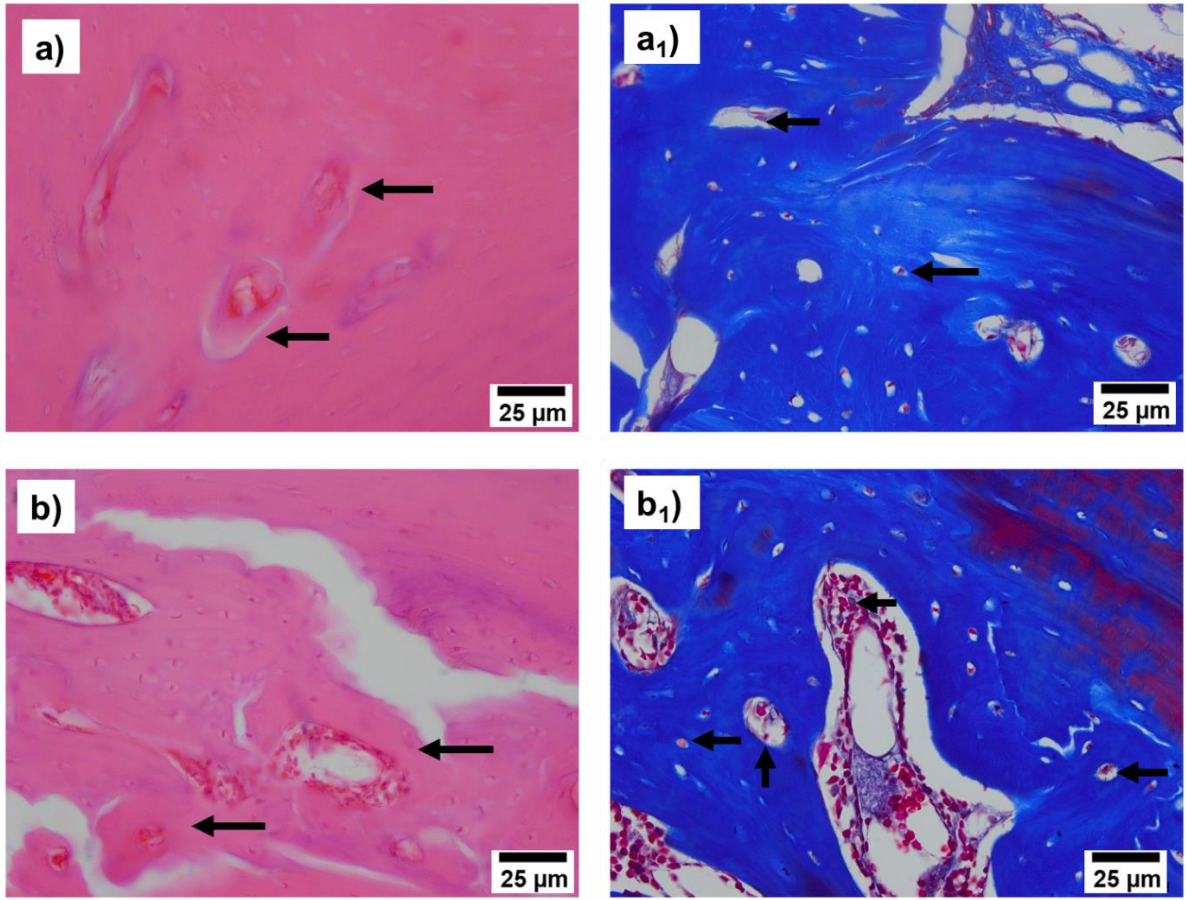

**Figure 7**. Micrographs of the defect filled with the G-nHAP scaffold, showing cranial vault bones. (a), (a₁) Control group without the G-nHAP scaffold. (b), (b₁) The defect filled with the G-nHAP scaffold. The bones were stained with Masson trichrome (a), (b), and Hemotoxin-eosin at the magnification of 400 times.

A reference Cerabone bone substitute shows similar structure and characteristics including a mature bone, mineralized bone, osteoids, osteocytes, and osteoprogenitor cells (Figure S11), suggesting that the G-nHAP scaffold is promising for future applications as a scaffold. The presence of the red color indicates bone hardness, agreeing with the high compressive strength of the G-nHAP. This can be assigned to the exceptional strength graphene layers that have a role in interconnecting the nHAP crystals, forming a compact bone structure.

## 4. Conclusions

Designing mechanically robust and compact bone scaffolds is an important challenge for resolving bone tissue regeneration in the field of biomedical engineering. The role of graphene nanowindows in the formation of nHAP scaffold during the hydrothermal synthesis process is introduced for the first time. The graphene layers were bridged through the nanowindows, giving the new structure of a tightly bound graphene layer with nHAP grown on graphene. The crystalline growth of the nHAP was suppressed at the crystallographic faces due to the tight G-nHAP scaffold formation that is promising for future fracture-resistive bone formation. The mechanical strength of the G-nHAP scaffolds was as high as $12.94 \pm 1.47$ MPa, being comparable to the cancellous bone. *In vivo* studies demonstrated the possibility of real application of the G-nHAP scaffold. After six weeks the mineralized bone and mature bone formed with visible osteoid for the bone formation. This study provides an overview of the scaffold from the preparation, understanding the structure and coupling graphene and nHAP through the nanowindows, and its successful use in Wistar rats. This synthesis method and a mechanism of G-nHAP formation will be promising for the future progress and implementation of graphene-based bone scaffolds.

**Acknowledgment**

This project has received funding from the European Union's Horizon 2020 research and innovation program under grant agreement No 101007417, having benefited from the access provided by Directorate General Joint Research Centre in Directorate F - Health and Food - Technologies for Health (F.2) within the framework of the NFFA-Europe Pilot Transnational Access Activity, proposal [ID448].